\documentclass[runningheads,onecollarge]{svjour2}       

\smartqed  
\usepackage{graphicx}
\usepackage{amsmath,amsfonts,bm} 
\usepackage{hyperref}
\hypersetup{
      colorlinks = true,
      citecolor = blue,
      linkcolor = blue,
      urlcolor = blue
}
\usepackage{latexsym}
\usepackage{bbding}
\begin{document}

\title{Nonequilibrium free energy methods applied to magnetic systems: the degenerate Ising model}

\titlerunning{Nonequilibrium free energy methods applied to magnetic systems}        

\author{Samuel Cajahuaringa and Alex Antonelli}

\authorrunning{S. Cajahuaringa and A. Antonelli} 

\institute{S. Cajahuaringa \at
samuelif@ifi.unicamp.br \\          
A. Antonelli \at
aantone@ifi.unicamp.br \\
\emph{Instituto de F\'{i}sica Gleb Wataghin, Caixa Postal 6165,\\
Universidade Estadual de Campinas, UNICAMP 13083-970, Campinas, S\~{a}o Paulo, Brazil}
}

\date{Received: date / Accepted: date}

\maketitle

\begin{abstract}
In this paper, we review the physical concepts of the nonequilibrium techniques for the calculation of free energies applied to magnetic systems using Monte Carlo simulations of different nonequilibrium processes. The methodology allows the calculation of the free energy difference between two different system Hamiltonians, as well as the free energy dependence on temperature and magnetic field for a given Hamiltonian. As an illustration of the effectiveness of this approach, we apply the methodologies to determine the phase diagram of a simple microscopic model, the degenerate Ising model. Our results show very good agreement with those obtained from analytical (theoretical) methods.

\end{abstract}

\section{Introduction}
Magnetic materials comprise a wide variety of substances, which are used in a large number of applications in our daily life, such as, storage data devices, personal computers, cell phones, credit cards, MRI machines, etc. It is hard to conceive a modern society without magnetic materials since they are becoming vital to its economic development\cite{BookCoey}. Magnetic materials exhibit a variety of properties when submitted to different external conditions. The knowledge of phase diagrams provides valuable information about phase changes and coexistence conditions, which is of great importance for the development of applications in different research areas.

The determination of phase diagrams by computer simulation requires the computation of the free energy of the various phases of the system, which is a notoriously difficult task. The difficulties stem from the fact that free energy depends on the volume in phase space available to the system, in contrast to other thermodynamic properties, such as internal energy, enthalpy, temperature, etc., that can be easily computed as averages of functions that depend explicitly on the phase space coordinates\cite{free}.

The thermodynamic integration (TI) method\cite{free} has been widely used to compute free energies, which consists on the construction of a sequence of equilibrium states along a path between two thermodynamic states of  interest, the calculated free energy difference between these two states is
essentially exact, provided that the system does not undergo a first order phase transition along the chosen thermodynamic path. Since the free energy of the reference system is known, the absolute free energy of the system of interest is readily obtained. In practice, since the integration along the thermodynamic path is performed numerically, it requires the calculation of several ensemble averages, which is computationally very demanding.

Free energy calculation methods have benefited substantially with the introduction of nonequilibrium approaches, such as the adiabatic switching (AS) method\cite{AS}, which allows the calculations to be performed along explicitly time-dependent processes. If such nonequlibrium processes connecting two equilibrium states are simulated sufficiently slowly, dissipation effects are small and one can, at least in principle, obtain the free energy of the system of interest performing only one nonequilibrium simulation\cite{AS3,isingfreeenergy}, which can lead to significant efficiency gains when compared to standard equilibrium methods.

Although the AS method is very efficient, it provides the free energy at a single thermodynamic state. However, if one is interested in knowing the free energy in a wide interval of temperatures, many simulations are required. This task can be efficiently done by the Reversible Scaling (RS) method\cite{RS}, which allows the calculation of the free energy for a wide interval of temperatures, requiring only the knowledge of the free energy at just a single temperature, which can be provided by the AS method. Recently, it was shown that the AS method allow the calculation of the free energy efficiently along an isothermal path\cite{Samuel}.

However, in order to compute phase boundaries still several AS and RS calculations are required. In order to optimize the calculation of phase boundaries, it was proposed by de Koning et al.\cite{CC} a dynamical integration of the Clausius-Clapeyron (dCCI) equation using nonequilibrium simulations. This method allows the calculation of the phase boundary using, in principle with a single simulation, whose length is comparable to a regular equilibrium simulation.

In this paper, we describe the state-of-the-art of nonequilibrium techniques for calculations of the free energies of magnetic systems. These methodologies were applied to obtain a phase diagram of the degenerate Ising model on a square lattice\cite{Harris,degenerate1}, since the analytical solution of the phase boundary to this system is known, we will be able to compare our numerical results to exact values.

The article is organized as follows. In Sect.~\ref{sec:Model} we review the thermodynamics and statistical mechanics of magnetic systems, particularly the degenerate Ising model. We also describe in Sect.~\ref{sec:Model} the Monte Carlo method to perform simulations at constant temperature and magnetic field. In Sect.~\ref{sec:FreeEnergy}, we present the methodology that, by using different nonequilibrium processes, allows to compute the free-energy difference between two different system Hamiltonians, as well as, the free energy dependence on temperature and magnetic field for a given Hamiltonian. Sect.~\ref{sec:FreeEnergy}, also provides the details of the dCCI methodology, in which from a given initial coexistence condition one is able to determine the entire phase boundary from a single nonequilibrium simulation. Sect.~\ref{sec:Results} gives the results of the methodologies applied to determine the phase boundary of the degenerate Ising model. Finally, in Sect.~\ref{sec:Conclusions} we summarize the main results of our work.
\section{Degenerate Ising Model}\label{sec:Model}
Let us consider the analogy between the gas-liquid and magnetic cases\cite{BookTuckerman}, the following thermodynamic variables, such as pressure and magnetic field ($P\Leftrightarrow B$) are equivalent, a similar relation holds for volume and magnetization ($V\Leftrightarrow-M$), thus all thermodynamic relations applicable to one case are still valid to the other one, when these equivalences are taken into account. The Gibbs free energy for a magnetic substance is given by
\begin{equation}
G(B,T)=U-BM-TS,
\label{eqn:thermodynamics_gibbs}
\end{equation}
where $U$ is the internal energy, which is related to the configurational energy of the system, $T$ is the temperature, $S$ the entropy of the system, and the term $BM$ is related to the magnetic work performed by the magnetic field. We can also recognize the term $U-BM$ as the magnetic enthalpy $E$.

An important model that was proposed in the context of magnetism, but turned out to bring important contributions to the understanding of fundamental physics, is the famous Ising model\cite{Baxter,Ising1}. It describes the behavior of classical magnetic moments that can point in two directions, either up or down, interacting via nearest-neighbor ferromagnetic or antiferromagnetic exchange, in a magnetic field, which is described by Hamiltonian
\begin{equation}
\mathcal{H}(\bm{\sigma})=-J\sum_{<i,j>}\sigma_{i}\sigma_{j}-B\sum_{i}\sigma_{i},
\label{eqn:H_ising}
\end{equation}
where $\bm{\sigma}=\{\sigma_{1},\sigma_{2},\ldots,\sigma_{N}\}$ represents a configuration of $N$ spins arranged in a $L\times L$ square lattice. Each spin has a value of $\pm 1$, and interacts only with its four nearest neighbors, as indicated by the summation over all nearest-neighbor pairs $<i,j>$. The interaction energy between the spins is $J>0$ for the ferromagnetic case and is impose periodic boundary conditions to avoid border effects.

Note that the Hamiltonian (Eq.~(\ref{eqn:H_ising})) does not correspond to the internal energy as it appears in many textbooks on elementary statistical mechanics and is strictly related to the magnetic enthalpy\cite{Castellano}. This point should be clear, since it allows us to make the correct association between the partition function and the Gibbs free energy, and also it permits to correctly apply the nonequilibrium free energy methods, which we will address later.

\begin{figure}[tbp]
     \centering
         \includegraphics[scale=0.44,bb=0 0 454 283,clip]{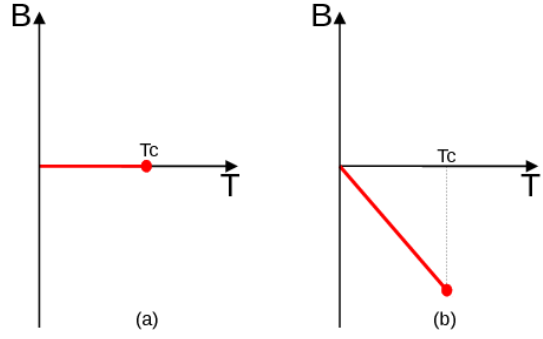}
     \caption{\label{fig:boundary-phase} The BT phase diagram for (a) the ferromagnetic Ising model and (b) the degenerate variant, carried to the spin-up. The red line represents a phase boundary that defines the first-order phase transition. The dot at $T=T_{c}$ represents the critical point. The slope of the boundary is $-\frac{1}{2}k_{B}\ln\delta$, and the value of $T_{c}$ is independent of $B$ or $\delta$.}
\end{figure}

The statistical mechanics of the spin-half Ising model in two dimensions in the absence of an external field $(B = 0)$  is well understood. There are a temperature-driven second-order transition at a specific temperature $T_{c}$ and a field-driven first-order transition at $B=0$ for all $T<T_{c}$ in the conventional Ising model (see Fig. \ref{fig:boundary-phase}a).

In order to study temperature-driven first-order transitions, Harris\cite{Harris} has modified the Ising model so that the first-order transition still occurs at the same temperature ($T<T_{c}$) but a particular magnetic field ($B\neq0$) by the addition of a parameter $\delta>0$, which adjusts the relative weightings of the two possible spin in the partition function. This weighting essentially introduces degenerate states into the system and $\delta$ represents the degeneracy ratio of spin-up to spin-down states. By varying $\delta$ one obtains an affine mapping of curves in the $(B,T)$ phase diagram: the value of $\delta$ completely determines the slope of the phase boundary in the BT plane, as exemplified in Fig. \ref{fig:boundary-phase}b.

The probability distribution function considering only the degeneracy of the spin-up is
\begin{eqnarray}
\rho(\bm{\sigma})&=&\mathcal{Z}^{-1}\delta^{N_{\uparrow}(\bm{\sigma})}e^{-\beta
E(\bm{\sigma})}  
\nonumber\\
&=&\mathcal{Z}^{-1}e^{-\beta(E(\bm{\sigma})-N_{\uparrow}(\bm{\sigma})k_{B}Tln\delta)},
\label{eqn:rho_sigma}
\end{eqnarray}
where $N_{\uparrow}(\bm{\sigma})$ is the number of spins with spin-up in the state $\bm{\sigma}$ and $E(\bm{\sigma})$ is the corresponding magnetic enthalpy. The partition function at constant temperature and magnetic field is
\begin{equation}
\mathcal{Z}=\sum_{i}\delta^{N_{\uparrow}(\bm{\sigma}_{i})}e^{-\beta  
E(\bm{\sigma}_{i})},
\end{equation}
the corresponding macroscopic thermodynamic potential is the Gibbs free energy,
\begin{equation}
G(B,T)=-k_{B}T\ln\mathcal{Z}.
\end{equation}

Note that the parameter $\delta$ in this case controls the degeneracy of the spin-up. Within the model proposed by Harris each Boltzmann factor contains an effective potential\cite{degenerate1}
\begin{eqnarray}
\tilde{E}(\bm{\sigma})\equiv E(\bm{\sigma})-N_{\uparrow}(\bm{\sigma})k_{B}T\ln\delta \qquad\\   
=-(B+\frac{1}{2}k_{B}T\ln\delta)\sum_{i}\sigma_{i}-J\sum_{<i,j>}\sigma_{i}\sigma_{j}-\frac{1}{2}Nk_{B}T\ln\delta \nonumber,
\end{eqnarray}
where the last term is a constant that is unimportant for the subjacent statistical mechanics and, in general, can be ignored. In fact, $\delta$ gives rise to a transformation,
\begin{equation}
B\mapsto B_{eff}=B+\frac{1}{2}k_{B}T\ln\delta,
\end{equation}
with the effective field ($B_{eff}$) going to zero at a specific temperature
\begin{equation}
T=\frac{-2B}{k_{B}\ln\delta}.
\label{eqn:coex_point}
\end{equation}
From these results, the slope of the boundary is $-\frac{1}{2}k_{B}\ln\delta$, thus determining the phase boundary of the degenerate Ising model.
\subsection{Metropolis scheme}
In this work, we perform Monte Carlo simulations of the degenerate Ising model, by using the Metropolis algorithm\cite{free}, where the equilibrium behavior under particular physical constraints is reached if the states are generated by the following acceptance probability
\begin{eqnarray}
w_{ij}&=&\min\{1,\rho(\bm{\sigma}_{i})/\rho(\bm{\sigma}_{j})\},\\
\rho(\bm{\sigma}_{j})&=&\mathcal{Z}^{-1}\delta^{N_{\uparrow}(\bm{\sigma}_{j})}e^{-\beta E(\bm{\sigma}_{j})}, \\
\rho(\bm{\sigma}_{i})&=&\mathcal{Z}^{-1}\delta^{N_{\uparrow}(\bm{\sigma}_{i})}e^{-\beta E(\bm{\sigma}_{i})},
\end{eqnarray}
where $\rho(\bm{\sigma}_{j})$ and $\rho(\bm{\sigma}_{i})$ are the probability distribution function of the current and proposed configuration states, respectively. $w_{ij}$ is the acceptance probability of the proposed state. Due to the degeneracy of the spin-up, it is necessary to analyze the acceptance probability of a single-flip separately.

If one picks a spin with spin-up and flips it to spin-down, the number of spins with spin-up is reduced by one
\begin{equation}
N_{\uparrow}(\bm{\sigma}_{i})=N_{\uparrow}(\bm{\sigma}_{j})-1,
\end{equation}
then the acceptance probability in this case is
\begin{equation}
w_{ij}=\min\{1,\delta^{-1}e^{-\beta(E(\bm{\sigma}_{i})-E(\bm{\sigma}_{j}))}\},
\end{equation}
else, if one picks a spin with spin-down and flips it to spin-up, the number of spins with spin-up is increased by one
\begin{equation}
N_{\uparrow}(\bm{\sigma}_{i})=N_{\uparrow}(\bm{\sigma}_{j})+1,
\end{equation}
then the acceptance probability in this case is
\begin{equation}
w_{ij}=\min\{1,\delta e^{-\beta(E(\bm{\sigma}_{i})-E(\bm{\sigma}_{j}))}\},
\end{equation}

Note that the criteria for single-flip the spin is not symmetric and can be summarized as
\begin{equation}
w_{ij} =
   \begin{cases}
    \min\{1,\delta^{-1}e^{-\beta(E(\bm{\sigma}_{i})-E(\bm{\sigma}_{j}))}\}  
& \uparrow\Rightarrow\downarrow \\
   \min\{1,\delta e^{-\beta(E(\bm{\sigma}_{i})-E(\bm{\sigma}_{j}))}\}  
& \downarrow\Rightarrow\uparrow.
   \end{cases}
\label{eqn:mc_weights}
\end{equation}
By setting $\delta=1$ it yields the usual Ising model with symmetric acceptance probability.
\section{Nonequilibrium free energy methods}\label{sec:FreeEnergy}
We now briefly discuss the basics of the methodology we use to compute the free energy. Since it is a thermal quantity that cannot be expressed in terms of an ensemble average, it cannot be computed directly using Monte Carlo (MC) or Molecular Dynamics (MD) sampling methods. As a result, free energies are usually determined using indirect strategies, in which free energy differences between two systems can be computed by evaluating the work associated with a reversible process that connects two systems\cite{free}.

In the TI method, free energy differences are computed by using the calculation of reversible work. The main idea behind the method is to consider a reference system, for which the free energy is known, and perform a connection to the system of interest by means of a varying coupling parameter $\lambda$. A typical functional form of this coupling is given by the Hamiltonian
\begin{equation}
\mathcal{H}(\lambda)=(1-\lambda)\mathcal{H}_{sys}+\lambda\mathcal{H}_{ref},
\label{eqn:H_coupled}
\end{equation}
where $\mathcal{H}_{sys}$ and $\mathcal{H}_{ref}$ represent the Hamiltonian of the system of interest and the reference, respectively. Note that this form allows a continuous switching between $\mathcal{H}_{ref}$ and $\mathcal{H}_{sys}$ by varying the parameter $\lambda$ between 0 and 1. It is straightforward to show, starting from the definition of the partition function, that the Gibbs free energy difference between the two systems is given by the reversible work ($W_{rev}$)
\begin{equation}
\Delta  
G=G_{ref}-G_{syst}=\int_{0}^{1}d\lambda\bigg\langle\frac{\partial\mathcal{H}(\lambda)}{\partial\lambda}\bigg\rangle_{\lambda}
\equiv  
W_{rev},
\label{eqn:work_rev}
\end{equation}
where the brackets indicate an equilibrium average in a statistical ensemble of interested. This integration gives the total work done by the generalized force $\partial\mathcal{H}/\partial\lambda$. Since it involves equilibrium averages of the system at all times, it reflects a reversible process. While the TI method is exact in principle, it requires several equilibrium simulations (at least one for each value of $\lambda$) to obtain accurate results.

In the nonequilibrium approach, the AS method estimates the integral of Eq.~(\ref{eqn:work_rev}) in terms of the irreversible work done along a single simulation in which $\lambda$ is explicitly time-dependent and varies from $\lambda(0)=0$ to $\lambda(t_{s})=1$ ($t_{s}$ is the total duration of the switching process)
\begin{equation}
W_{irr}=\int_{0}^{t_{s}}dt'\frac{d\lambda}{dt}\bigg|_{t'}\frac{\partial\mathcal{H(\lambda)}}{\partial\lambda}\bigg|_{\lambda(t')}.
\label{eqn:Wirr_AS}
\end{equation}

Given the intrinsic irreversible nature of the process, this dynamic estimator is biased, subject to both statistical and systematic errors. The statistical errors are due  to the fact that $W_{irr}$ is a stochastic quantity that depends on the initial conditions, while the systematic error is associated with the dissipative entropy production ($Q\geq0$) characteristic of irreversible processes. Dissipation effects, however, can be controlled by the simulation time and how the coupling parameter varies with time. In this case, within the linear response approximation, the systematic error is independent of the switching process direction, i.e., the entropy production is equal for the forward and backward, processes\cite{AS3,CC}. In this way, we can obtain an unbiased estimate for $W_{rev}$ according to
\begin{equation}
W_{rev}=\frac{1}{2}\big(\overline{W}_{irr}^{\lambda:0\to1}-\overline{W}_{irr}^{\lambda:1\to0}\big)\label{eqn:W_irr}
\end{equation}
where $\overline{W}_{irr}$ is the average of the $W_{irr}$ in an ensemble of nonequilibrium simulations with different initial conditions for a given switching time $t_{s}$. Similarly, the magnitude of the dissipation in this regime can be estimated by
\begin{equation}
Q=\frac{1}{2}\big(\overline{W}_{irr}^{\lambda:0\to1}+\overline{W}_{irr}^{\lambda:1\to0}\big)\label{eqn:Q_dissp}
\end{equation}

To implement the AS method in Monte Carlo simulations, the acceptance probabilities need to depend explicitly on the value of the coupling parameter $\lambda$ along the nonequilibirum processes
\begin{equation}
w_{ij}(\lambda) =
   \begin{cases}
\min\{1,\delta^{-1}e^{-\beta(E(\bm{\sigma}_{i};\lambda)-E(\bm{\sigma}_{j};\lambda))}\}
      &  \uparrow\Rightarrow\downarrow\\
     \min\{1,\delta  e^{-\beta(E(\bm{\sigma}_{i};\lambda)-E(\bm{\sigma}_{j};\lambda))}\}  & \downarrow\Rightarrow\uparrow,
   \end{cases}
\end{equation}
and $E(\bm{\sigma};\lambda)$ is the energy at a specific coupling parameter
\begin{equation}
E(\bm{\sigma};\lambda)=(1-\lambda)E_{sys}(\bm{\sigma})+\lambda  
E_{ref}(\bm{\sigma}),
\end{equation}
where $E_{sys}(\bm{\sigma})$ and $E_{ref}(\bm{\sigma})$ are the energies of the system of interest and of reference, respectively, for a specific spin configurations ($\bm{\sigma}$).

The simulation consists of two parts, first a MC sweep is performed through the entire system using the acceptance probabilities corresponding to the current value of $\lambda$. After this sweep, the values of coupling parameter $\lambda$ and the corresponding acceptance probability are updated for generating the next system configuration.

Next, we discuss the application of the methodology in the calculation of the free energy difference between two systems described by different Hamiltonians and the computation of the free energy along a wide interval of temperature and magnetic field.
\subsection{Non-interacting spins as a reference system}\label{sec:AS_ref}
The choice of a suitable reference depends on the nature of the system of interest. For the computation of the free energy of the degenerate Ising model at high temperature, where is expected to observe the paramagnetic phase, we use the non-interacting spins system with same degeneracy of the degenerate Ising model as the natural reference system, since its free energy is known analytically and it is always possible to construct a reversible path to connect both systems. The Hamiltonian of the reference system is
\begin{equation}
\mathcal{H}_{ref}(\bm{\sigma})=-B\sum_{i}\sigma_{i}
\label{eqn:H_ref}
\end{equation}
and the Gibbs free energy is
\begin{equation}
G_{ref}(N,B,T)=-Nk_{B}T\ln\bigg(e^{-\frac{B}{k_{B}T}}+\delta e^{\frac{B}{k_{B}T}}\bigg).
\label{eqn:G_ref}
\end{equation}

By applying the AS method to estimate the $W_{rev}$ between the two systems (Eq.~\ref{eqn:W_irr}), the free energy of the system of interest can be estimated as
\begin{equation}
G_{sys}(N,B,T)=W_{rev}+G_{ref}(N,B,T).
\end{equation}
\subsection{Temperature dependence of the free energy: the reversible scaling paths}\label{sec:RS_T}
The RS method is very efficient to obtain the free energy in a large interval of temperatures\cite{RS}. It is based on an equivalence relation between the partition functions of a given system of interest and its associated scaled systems. The free energy of the scaled system can be easily determined by the AS method. Therefore, one can, at least in principle, determine the free energy in a wide interval of temperatures from only a single nonequilibrium simulation, whose length is similar to those used to obtain thermodynamical quantities that are simple averages in phase space, such as energy, enthalpy, etc. In order to facilitate the derivation of the dCCI method for magnetic systems, in this section we present the key equations of this method which was originally described in Refs\cite{RS,CC}.

Let us consider the Harris's model at a constant temperature $T$ and magnetic field $B$, the Gibbs free energy of the system of interested is given by
\begin{equation}
G(B,T)=-k_{B}T\ln  
\Big[\sum_{i}\delta^{N_{\uparrow}(\mathbb{\sigma}_{i})}e^{-\frac{1}{k_{B}T}\big(-J\sum\limits_{<i,j>}\sigma_{i}\sigma_{j}-B\sum\limits_{i}\sigma_{i}\big)}\Big].
\label{eqn:Gibbs_system}
\end{equation}

Consider now the Gibbs free energy of a system whose configurational energy is scaled by $\lambda$ at a constant temperature $T_{0}$ and magnetic field $B_{s}(\lambda)$, we have
\begin{equation}
G_{s}(B_{s}(\lambda),T_{0})=-k_{B}T_{0}\ln\Big[\sum_{i}\delta^{N_{\uparrow}(\mathbb{\sigma}_{i})}e^{-\frac{1}{k_{B}T_{0}}\big(-\lambda J\sum\limits_{<i,j>}\sigma_{i}\sigma_{j}-B_{s}(\lambda)\sum\limits_{i}\sigma_{i}\big)}\Big].
\label{eqn:Gibbs_scaled}
\end{equation}

Using the scaling relations
\begin{equation}
T=\frac{T_{0}}{\lambda}
\end{equation}
and
\begin{equation}
B_{s}(\lambda)=\lambda B,
\end{equation}
we combine Eqs.~(\ref{eqn:Gibbs_system}) and (\ref{eqn:Gibbs_scaled}) resulting in
\begin{equation}
\frac{G(B,T)}{T}=\frac{G_{s}(B_{s}(\lambda),T_{0})}{T_{0}}.
\label{eqn:RS_Gibbs}
\end{equation}

This relation implies that the problem of computing the Gibbs free energy $G(T,B)$ of the system of interest is completely equivalent to the problem of evaluating the Gibbs free energy of the scaled system at temperature $T_{0}$, provided that the potential energy and magnetic field of the latter system are those of the former one properly scaled by a factor $\lambda$. The importance of this result is that if we know the Gibbs free energy of the scaled system for a given reference value $\lambda_{ref}$, the Gibbs free energy for any value of $\lambda$ is given by
\begin{equation}
G_{s}(B_{s}(\lambda),T_{0})=G_{s}(B_{s}(\lambda_{ref}),T_{0})+W_{rev}(\lambda),
\end{equation}
where
\begin{equation}
W_{rev}(\lambda)=\int_{\lambda_{ref}}^{\lambda}d\lambda'\Big[\langle  
U\rangle-\frac{dB_{s}}{d\lambda}\Big|_{\lambda'}\langle M\rangle\Big],
\label{eqn:W_rev_scaled}
\end{equation}
is the reversible work function associated with the scaling coordinate $\lambda$. Here, $\langle U\rangle$ and $\langle M\rangle$ denote ensemble averages of the configurational energy and magnetization, respectively, computed for the scaled system at temperature $T_{0}$, scaling parameter $\lambda$ and magnetic field $B_{s}(\lambda)$.

The reversible work in Eq.~(\ref{eqn:W_rev_scaled}) can be obtained within very good accuracy through the irreversible work done when $\lambda(t)$ is varied from $\lambda(0)=\lambda_{ref}$ to $\lambda(t_{s})=\lambda$ during a single simulation performed at temperature $T_{0}$:
\begin{equation}
W_{irr}(\lambda(t))=\int_{0}^{t_{s}}dt\frac{d\lambda(t)}{dt}\Big[U(t)-\frac{dB_{s}}{d\lambda}\Big|_{\lambda(t)}M(t)\Big].
\label{eqn:W_irr_scaled}
\end{equation}

Therefore, we can write Eq.~(\ref{eqn:RS_Gibbs}) as
\begin{equation}
\frac{G(B,T)}{T}=\frac{G_{s}(B_{s}(\lambda_{ref}),T_{0})}{T_{0}}+\frac{1}{T_{0}}\int_{0}^{t_{s}}dt\frac{d\lambda(t)}{dt}\Big[U(t)-\frac{dB_{s}}{d\lambda}\Big|_{\lambda(t)}M(t)\Big].
\label{eqn:Gibbs_RS_T}
\end{equation}

Thus, for a given magnetic field $B$, Eq.~(\ref{eqn:Gibbs_RS_T}) allows one to obtain the Gibbs free energy of a system in a wide interval of temperatures using a single nonequilibrium simulation, provided that the Gibbs free energy of the scaled system at a reference state is known.
\subsection{Magnetic field dependence of the free energy: the adiabatic switching paths}\label{sec:AS_H}
The AS method is commonly used to provide the free energy of a single point, where only the initial and final points on the trajectory correspond to physical relevant systems. In this article, we review the formulation of the AS method in order to determine the Gibbs free energies over a wide magnetic field interval from a single simulation. Let us consider a system at different magnetic fields, but at the same temperature. It is possible to compute the difference of the Gibbs free energies at these two different magnetic fields by evaluating the work associated with a reversible process along a path connecting the physical system at the reference magnetic field $B_{0}$ to the system at a magnetic field of interest $B$. The corresponding Gibbs free energies are then given by
\begin{equation}
G(B_{0},T)=-k_{B}T\ln\Big[\sum_{i}\delta^{N_{\uparrow}(\bm{\sigma}_{i})}e^{-\frac{1}{k_{B}T}(-J\sum\limits_{<i,j>}\sigma_{i}\sigma_{j}-B_{0}\sum\limits_{i}\sigma_{i})}\Big],
\label{eqn:Gibbs_B0}
\end{equation}
is the Gibbs free energy of reference, and
\begin{equation}
G(B,T)=-k_{B}T\ln\Big[\sum_{i}\delta^{N_{\uparrow}(\bm{\sigma}_{i})}e^{-\frac{1}{k_{B}T}(-J\sum\limits_{<i,j>}\sigma_{i}\sigma_{j}-B\sum\limits_{i}\sigma_{i})}\Big],
\label{eqn:Gibbs_B}
\end{equation}
is the Gibbs free energy at the magnetic field of interest. By taking the derivative with respect to the magnetic field $B$ in Eq.~\ref{eqn:Gibbs_B}, one finds the equilibrium average of the magnetization at constant temperature and magnetic field. By integrating this result with respect to the magnetic filed, between the $B_{0}$ and $B$ at constant temperature, one obtains the Gibbs free energy difference along the isothermal path (TI method), which is given by
\begin{equation}
G(B,T)-G(B_{0},T)=W_{mech}=-\int_{B_{0}}^{B}dB'\langle M\rangle,
\label{eqn:W_mech}
\end{equation}
this integration gives the reversible mechanical work $W_{mech}$ done by the applied the external magnetic field, we also can estimate that in terms of the irreversible work $W_{irr}$, as determined by the RS method by single nonequilibrium simulation during which the value of magnetic field $B(t)$ changes dynamically, in such a way that, at the beginning of the simulation $B(0)=B_{0}$ and at the end $B(t_{s})=B$
\begin{equation}
W_{irr}=-\int_{0}^{t_{s}}dt'\frac{dB(t)}{dt}\bigg|_{t'}M(t'),
\label{eqn:W_mech_irr}
\end{equation}
as a consequence, one obtains the Gibbs free energy of a system in a wide interval of the magnetic field using one single simulation, provided that the Gibbs free energy of the system of interest in a reference state is known.
\subsection{Dynamic Clausius-Clapeyron integration}\label{sec:d_CCI}
The Clausius-Clapeyron equation is one of the most important results of thermodynamics, being regarded as perhaps the first successful application of thermodynamics to a physical problem\cite{Pippard}. Its derivation is based upon the constraint that at any point of the coexistence line the molar Gibbs free energy of both phases should be the same. For magnetic systems, the Clausius-Clapeyron equation gives the slope of the phase boundary in the $B,T$ plane
\begin{equation}
\frac{dB}{dT}=-\frac{\Delta H}{T\Delta M},
\label{eqn:CCE}
\end{equation}
where $\Delta H$ and $\Delta M$ are the molar magnetic enthalpy and magnetization differences between the two phases, respectively. Therefore, if a point of the coexistence line is known, one can obtain the whole phase boundary from Eq. \ref{eqn:CCE}.

In order to avoid performing a series of independent equilibrium simulations, de Koning et al.\cite{CC} based on an idea similar to the RS method, proposed a procedure by which the integration of the Clausius-Clapeyron equation is performed dynamically. Starting at a single point of the coexistence curve, this technique allows the exploration of the coexistence curve over a wide range of states using a single nonequilibrium simulation\cite{CC}. Here, we derive the dynamic Clausius-Clapeyron integration technique to magnetic systems.

Thus, starting from a known condition of phase coexistence, considering both phases $a$ and $b$, subject to reversible perturbations $d\lambda$ and $d\lambda(dB_{s}/d\lambda)$, the free energy of the scaled systems will change according to
\begin{equation}
dG_{s,a}=dW_{rev,a}=d\lambda\Big[\langle  
U\rangle_{a}-\frac{dB_{s}}{d\lambda}\langle M\rangle_{a}\Big]
\end{equation}
and
\begin{equation}
dG_{s,b}=dW_{rev,b}=d\lambda\Big[\langle  
U\rangle_{b}-\frac{dB_{s}}{d\lambda}\langle M\rangle_{b}\Big].
\end{equation}

In order to keep the phase-coexistence upon the application of this disturbance, we must have that $dG_{s,a}=dG_{s,b}$. Therefore,
\begin{equation}
\frac{dB_{s}}{d\lambda}=\frac{\langle U\rangle_{a}-\langle  
U\rangle_{b}}{\langle M\rangle_{a}-\langle M\rangle_{b}}.
\label{eqn:CCI_scaling}
\end{equation}

Similarly to what is done to estimate the free energy from nonequilibrium simulations, we can change Eq. \ref{eqn:CCI_scaling} in a way that the parameter $\lambda$ is varied dynamically, transforming it into the dynamic Clausius-Clapeyron equation
\begin{equation}
\frac{dB_{s}}{dt}=\dot{\lambda}(t)\frac{U_{a}(t)-U_{b}(t)}{M_{a}(t)-M_{b}(t)},
\label{eqn:d_CCI}
\end{equation}
where $\dot{\lambda}(t)$ is the rate of change of the scaling parameter and the ensemble averages of the potential energies and magnetization have been replaced by instantaneous values along the time-dependent process. Provided that the dynamic process is ideally reversible, the coexistence curve is given by
\begin{equation}
B_{s}^{\lambda(t)}=B_{s}^{\lambda(0)}+\int_{0}^{t}dt'\dot{\lambda}(t')\frac{U_{a}(t')-U_{b}(t')}{M_{a}(t')-M_{b}(t')}.
\end{equation}

Given an initial coexistence condition, the integration of this equation then provides a dynamical estimator for the entire coexistence curve from a single RS-AS simulation in which both phases are considered simultaneously.
\section{Coexistence curve for degenerate Ising model}\label{sec:Results}
As an illustration of the effectiveness of the nonequilibrium free energy methods applies to magnetic systems, we determined the coexistence boundary of the Ising model on a square lattice with degenerate states. The results presented in this section are obtained using Metropolis-like weights given in Eq.~(\ref{eqn:mc_weights}) and distribution $\rho(\mathbb{\sigma})$ defined in Eq.~(\ref{eqn:rho_sigma}). For convenience, from now on we adopt the following dimensionless quantities: $T\equiv k_{B}T/J$, $B\equiv B/J$, $M\equiv M/NJ$ and $G\equiv G/NJ$, where $T$, $B$, $M$ and $G$ are the temperature, magnetic field, magnetization and Gibbs free energy, respectively.
\subsection{Magnetization cycles}
\begin{figure}[bp]
     \centering
         \includegraphics[scale=0.38,bb=20 41 731 544,clip]{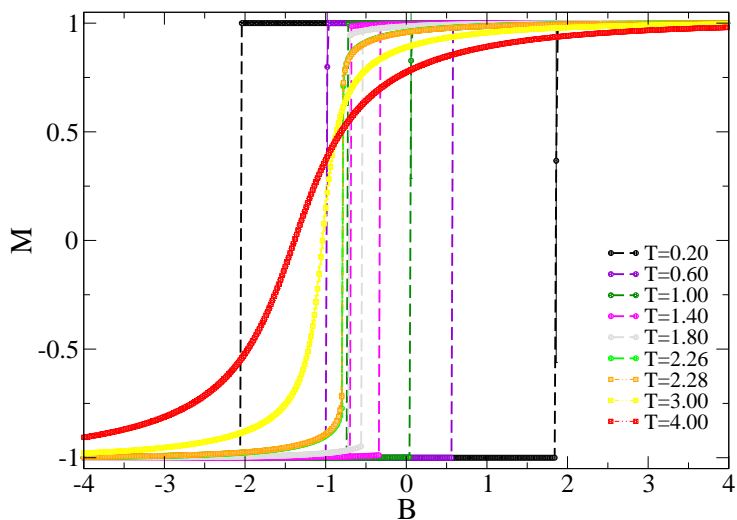}
     \caption{\label{fig:hysteresis} Cycles of the magnetization per spin as a function of the applied magnetic field for the degenerate Ising model on $100\times100$ square lattice (with degeneracy parameter $\delta=2$).}
\end{figure}

First, we explored the phase diagram surrounding the phase transition region between the spin-up and spin-down phases by performing simulations on a square lattice $(100\times100)$ with degeneracy parameter $\delta=2$, The cycles of magnetization as a function of magnetic field at a fixed temperature were obtained by ramping the magnetic field in increments of $0.02$. At each increment, the system was first equilibrated for $10^{4}$ MC sweeps and the averages were accumulated over another $10^{4}$ MC sweeps. These results are presented in Fig. \ref{fig:hysteresis}.
%

In Fig. \ref{fig:hysteresis} one can observe that the magnetization changes continuously for $T>T_{c}\approx2.27$, consistent with the transition being second-order and for temperatures $T<T_{c}$ the presence of hysteresis verifies that the transition is first-order. Note that in this case, the magnetization cycles are not symmetrical around zero magnetic field, in other words, the boundary phase that separates the spin-up and spin-down phases do not occur only at at zero magnetic fields, as it is expected, due to the degeneracy of the spin-up phase in this model.
\subsection{Path protocol}
\begin{figure}[tbp]
     \centering
          \includegraphics[scale=0.44,bb=37 6 668 519,clip]{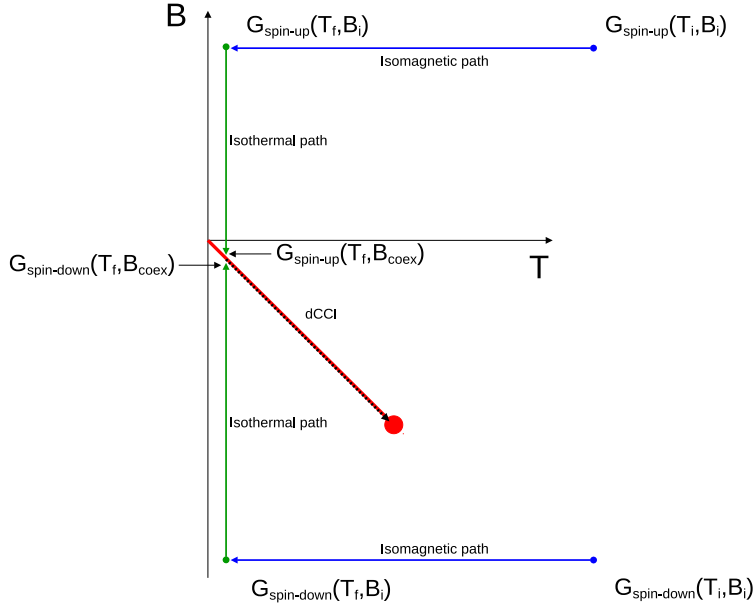}     
     \caption{\label{fig:protocol} Protocol path in order to determine the boundary phase through the calculation of the free energies. The blue points are the Gibbs free energies of the paramagnetic phase. The blue arrows indicated the calculation of the Gibbs free energy using the RS method along the isomagnetic path. The green points are the Gibbs free energies of the ferromagnetic phases. The green arrows indicated the calculation of the Gibbs free energy by the AS method along the isothermal path. After the crossing of the two Gibbs free energies the dCCI method is applied, which is indicated by the black dashed arrow.}
\end{figure}

In order to determine the boundary phase through the calculation of the free energy using nonequilibrium simulations, we performed the following protocol as shown in the Fig. \ref{fig:protocol}.

Initially, we determined the absolute value of the Gibbs free energy of the paramagnetic phase at two different magnetic fields and same temperature above the critical temperature ($T>T_{c}$) by using the AS method, this requires a reference system, whose Gibbs free energy is known in advance, we have chosen the non-interacting spin model with degeneracy as a reference system.

Using these Gibbs free energies computed for the paramagnetic phase at different magnetic fields, we apply the RS method at constant magnetic field to calculate the Gibbs free energy as a function of temperature. The system is cooled down below $T_{c}$ in order to obtain the Gibbs free energies of the ferromagnetic phases at the two different magnetic fields. After that, starting from these two states in the ferromagnetic phase,  we employ the AS method along the isothermal path, in order to determine the Gibbs free energies of the ferromagnetic phase as a function of the magnetic field. From the crossing of the two Gibbs free energy curves, we are able to determine the coexistence point. From this result, we can now apply the dCCI method to calculate the coexistence curve of this model.
\subsection{Gibbs free energy convergence tests}
\begin{figure}[tbp]
     \centering
     \begin{minipage}[b]{0.4\linewidth}
         \hspace*{-2.0cm}
                  \includegraphics[scale=0.32,bb=9 57 722 521,clip]{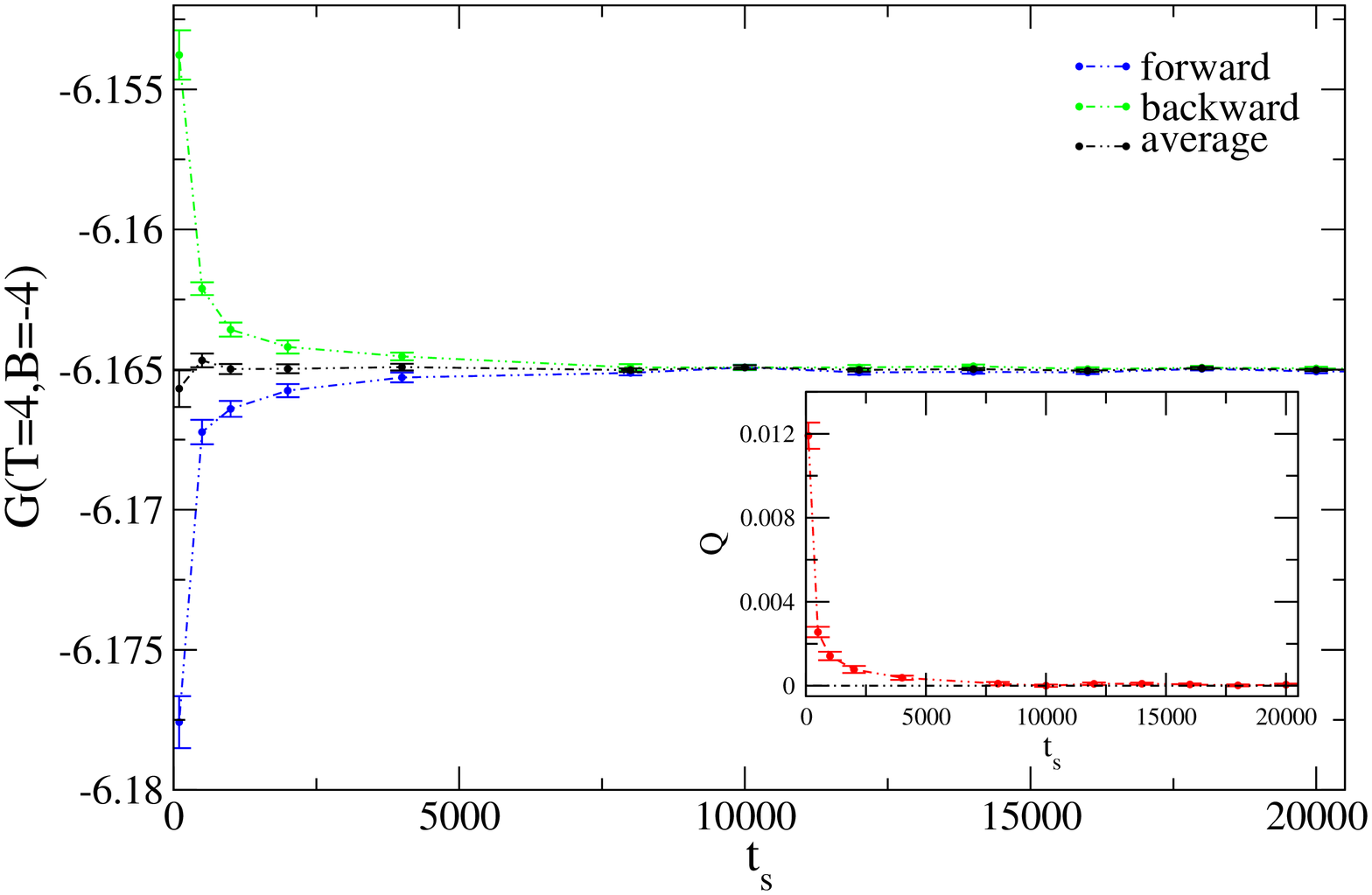}
     \end{minipage}
     \begin{minipage}[b]{0.4\linewidth}
         \includegraphics[scale=0.32,bb=9 57 722 521,clip]{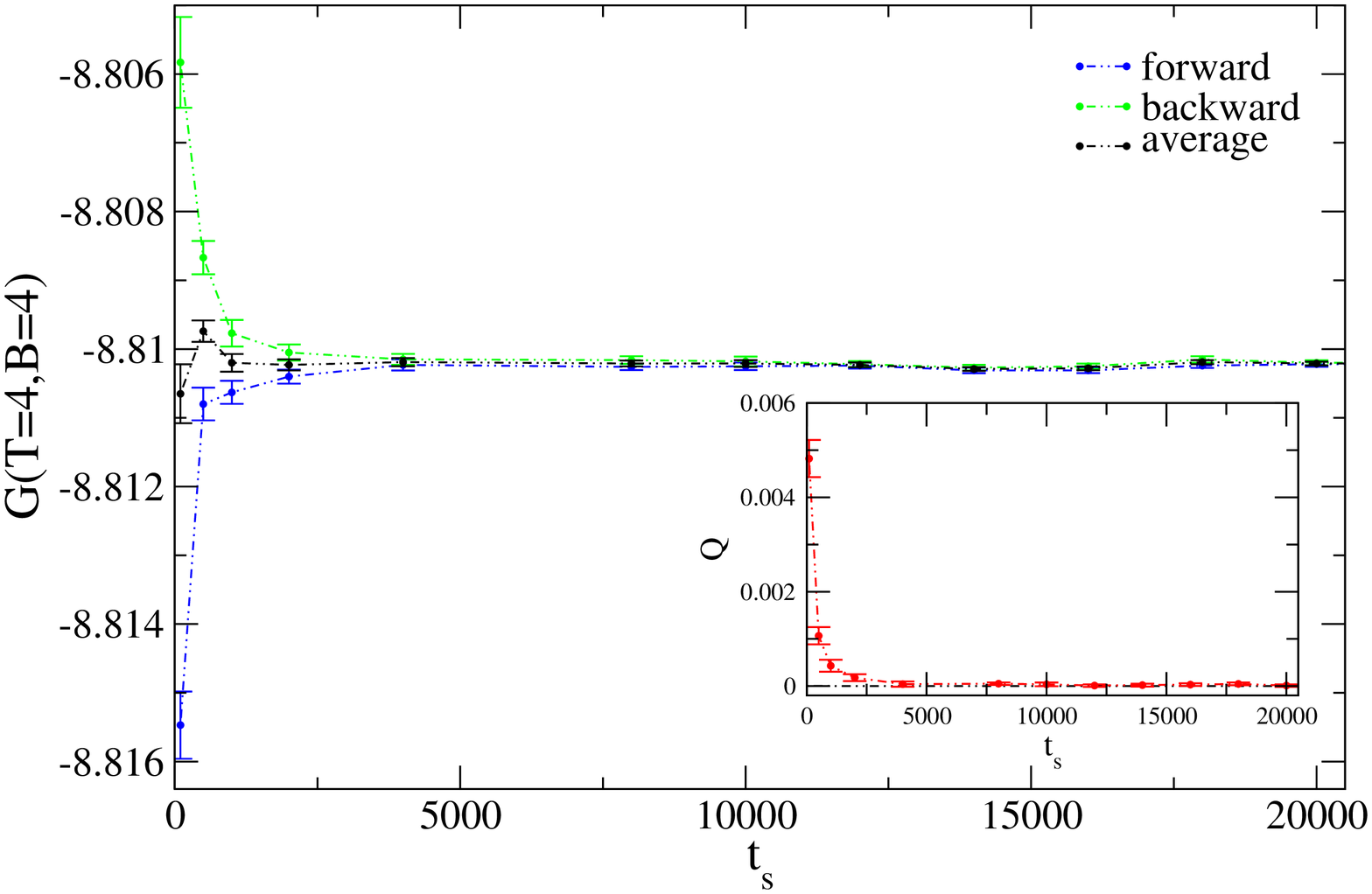}     
     \end{minipage}
     \caption{\label{fig:as_gibbs} Gibbs free energies calculations for the paramagnetic phases by the AS method as function of switched MC sweeps, at the temperature of $T=4.0$, the magnetic fields are $B=4.0$ (\textbf{left}) and $B=-0.4$ (\textbf{right}), respectively. The black, blue and green points correspond to average, the forward and backward estimator of the Gibbs free energy, respectively. (\textbf{Insets}) The dissipation energy in the switched process.}
\end{figure}

The graphs displayed in Fig. \ref{fig:as_gibbs} show the average, forward and backward, of the Gibbs free energy computed using the AS method of the paramagnetic phases at the temperature of $T=4.0$ for two different magnetic fields (Fig. \ref{fig:as_gibbs}(left) $B=4.0$ and Fig. \ref{fig:as_gibbs}(right) $B=-4.0$), twenty independent switching realizations were performed so that we could obtain an estimate for the statistical error.

The Fig. \ref{fig:as_gibbs} show how the computed Gibbs free energy converges with increasing the switching time $t_{s}$. In particular, we see that the average of the forward and backward paths is extremely efficient in eliminating the systematic error of the nonequilibrium approach, for $t_{s}>5000$ sweeps is sufficiently long for the linear response regime to be reached in both cases, since the dissipation energy is practically negligible (insets Fig.\ref{fig:as_gibbs}).

\begin{figure}[bp]
     \centering
         \includegraphics[scale=0.38,bb=24 63 716 521,clip]{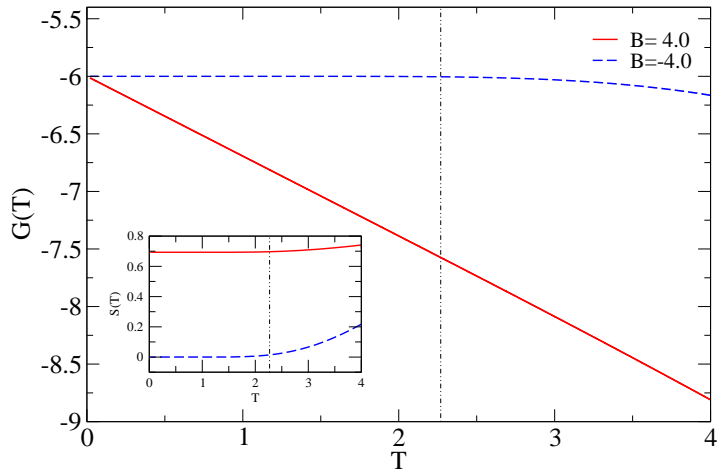}     
     \caption{\label{fig:RS_T} Gibbs free energy as a function of the temperature for two different magnetic fields, blue dashed line and red line corresponds to the $B=4$ and $B=-4$, respectively. (\textbf{Inset}) Entropy as a function of temperature for the two different magnetic fields.}
\end{figure}

Based on these results, we used the values of $G(T=4.0)$ for the degenerate non-interacting spins, which were computed at the two different values of the magnetic field ($B=4.0$ and $B=-0.4$), as the reference states to be used in the RS integration. The RS path chosen was such that $\lambda$ varies from 1 to 200, providing the Gibbs free energy temperature dependence from $T=4.0$ to $T=0.02$. In these simulations the magnetic field  was kept fixed at the values above specified. In Fig. \ref{fig:RS_T} we show well converged curves for the Gibbs free energy in the interval from $T=4$ to $T=0.02$, using a nonequilibrium switching as short as $5\times10^{4}$ MC sweeps.
Fig. \ref{fig:RS_T} also shows, in the inset, the entropy of both phases in the same interval of temperatures, calculated employing the thermodynamic relation (Eq.~(\ref{eqn:thermodynamics_gibbs})). As expected, the entropy of the spin-down phase close to zero temperature is null because the spin-down is non-degenerate, however our calculations show that the spin-up phase displays a residual entropy $S=0.693$, since the spin-up, in this case, is doubly degenerate, this result is in perfect agreement with the theoretical value of the entropy for the spin-up $S_{teor}=\ln2=0.6931$.
\subsection{Coexistence point}
\begin{figure}[tbp]
     \centering
         \includegraphics[scale=0.38,bb=39 60 716 529,clip]{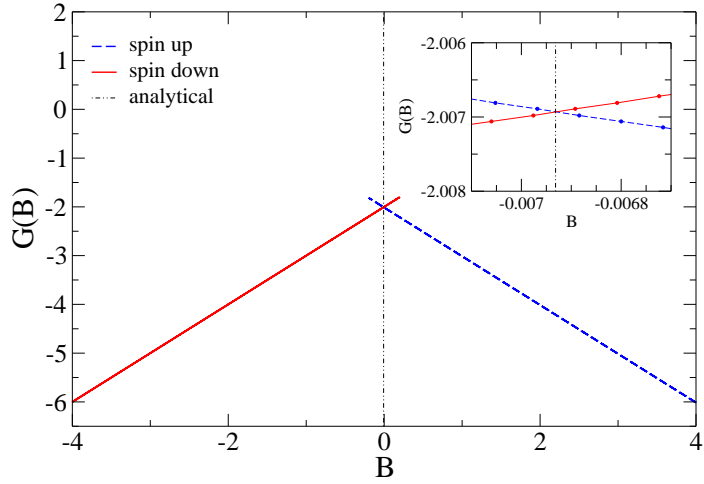}
     \caption{\label{fig:AS_H} Gibbs free energy as a function of the magnetic field for both ferromagnetic phases, blue dashed line and red line corresponds to the spin-up phase and spin-down phase, respectively and the black dashed line indicated the theoretical magnetic field for this crossing. (\textbf{Inset}) Gibbs free energy of the ferromagnetic phases around the theoretical magnetic field.}
\end{figure}
In order to determine the coexistence point for the ferromagnetic phases at low temperature ($T=0.02$), were perform AS integration along the isothermal paths (SubSec.~\ref{sec:AS_H}). We use the values of the Gibbs free energy for the two ferromagnetic phases as the reference point for the AS integration. For the ferromagnetic spin-up phase and spin-down phase, the magnetic field is changed dynamically for the $B=4.0$ to $B=-0.2$ and the $B=-4.0$ to $B=0.2$, respectively.
\begin{figure}[bp]
     \centering
         \includegraphics[scale=0.38,bb=22 57 701 529,clip]{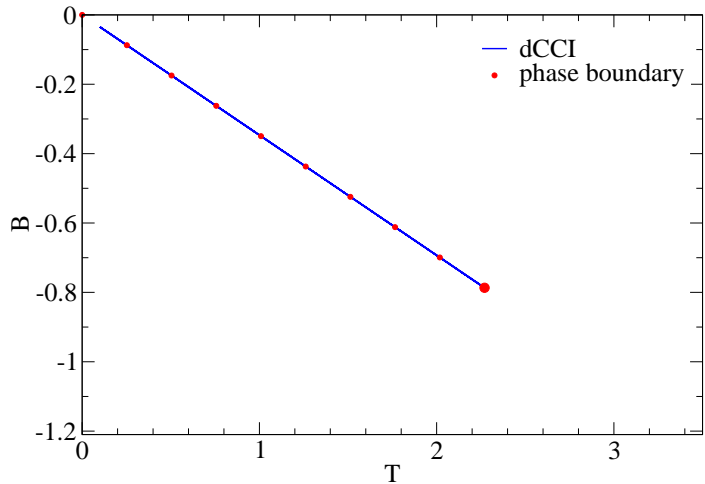}
     \caption{\label{fig:dCCI} Coexistence curve of the ferromagnetic phases for the Harris's model with double degeneracy of the spin-up in the B-T domain. The blue line is the result of the dCCI simulation, the circles are the theoretical value of this model.}
\end{figure}

In Fig. \ref{fig:AS_H} we show the crossing of the Gibbs free energy curves. Using a nonequilibrium switching as short as $5\times10^{4}$ MC sweeps, we are able to determine that at $T=0.02$ the crossing occurs at the magnetic field of $B_{coex}=-0.00693$, which is in perfect agreement with the theoretically predicted value from Eq.~(\ref{eqn:coex_point}), $\frac{1}{2}0.02\ln2=-0.00693147$. In the inset of the Fig. \ref{fig:AS_H}, we see that crossing of the Gibbs free energy curves occurs very close to the theoretical value indicate by the black dashed line.
\subsection{Coexistence curve}
Using the coexistence point as a starting point, we applied the dCCI method to extrapolate the coexistence curve of the ferromagnetic phases (below the critical temperature $T_{c}=2.27$). Our simulation used $10^{5}$ MC sweeps to reach the final magnetic field $B=-0.783257$ at the temperature $T=2.26$. The result is shown in Fig. \ref{fig:dCCI}, together with the theoretical value of the coexistence curve for this model with double degeneracy of the spin-up (Eq.~(\ref{eqn:coex_point})). We see in the Fig. \ref{fig:dCCI} the excellent agreement between the dCCI results and the analytical data obtained by the Eq.~(\ref{eqn:coex_point}). It is remarkable that even without carrying out an explicit estimation of systematic and statistical errors, the dCCI results agree very well with the exact solution, as can be seen by comparing the slope of the coexistence curve determined by the dCCI method, $-0.34657$. with the theoretical result, $-\frac{1}{2}\ln2=-0.3466$. We also determined the degeneracy of the spin-up from the dCCI results to be $\delta_{dCCI}=1.99999$.

The methodologies presented in this study constitute a very efficient and precise tool to study the phase diagram of magnetic systems.
\section{Conclusions}\label{sec:Conclusions}
In this work, we have illustrated how the nonequilibrium free energy methods are easily adapted to the study of the phase diagram of magnetic systems. Care has been taken in the derivation of thermodynamical properties of the magnetic system from basic statistical mechanics. The correct association of the partition function with Gibbs free energy has been achieved, allowing us the correct application of Monte Carlo simulation and the nonequilibrium free energy methods.

The methodology was applied to investigate the phase boundary of a representative example, the degenerate Ising model, which coexistence line can be readily evaluated analytically, our numerical results show that this coexistence line can be obtained, within a very good degree of accuracy, from a simulation comprising only $10^{5}$ MC sweeps, which is equivalent to the length of a regular equilibrium simulation.

The methodology is very competitive since it can compute the free energies and phase boundary of magnetic systems, for wide intervals of temperature and magnetic field, at a computational cost comparable to equilibrium simulations. This work adds to the effort of making nonequilibrium free energy methods accessible to a wider audience.

\begin{acknowledgements}
We gratefully acknowledge support from the Brazilian agencies CNPq, CAPES, and FAPESP under Grants \#2010/16970-0 and \#2016/23891-6. The calculations were performed at CCJDR-IFGW-UNICAMP and at CENAPAD-SP in Brazil.
\end{acknowledgements}

\bibliographystyle{spmpsci}      

\end{document}